\documentstyle[aasms4,12pt]{article}
\begin{document}

\title{WHAT CONTROLS THE STAR FORMATION IN LUMINOUS STARBURST MERGERS ?}

\author{Yoshiaki Taniguchi and Youichi Ohyama}

\affil{Astronomical Institute, Tohoku University, Aoba, Sendai 980-8578, Japan}


\begin{abstract}
In order to understand  what controls the star formation process in 
luminous starburst mergers (e.g., NGC 6240, Arp 220, and so on),
we investigate observational properties of two samples of high-luminosity 
starburst galaxies mapped 
in CO($J$=1--0) independently using both the Owens Valley Radio Observatory 
(Scoville et al. 1991) and the IRAM interferometer (Downes \& Solomon 1998). 
We find that the surface density of far-infrared luminosity, $\Sigma$(FIR),
is proportional linearly to the H$_2$ surface mass density, $\Sigma$(H$_2$),
for the two samples;  
$\Sigma$(FIR) $\propto \Sigma$(H$_2$)$^{1.01\pm0.06}$
with a correlation coefficient of 0.96.
It is often considered that $\Sigma$(FIR) provides a good measure of 
the star formation rate per unit area, $\Sigma$(SFR).
It is also known that molecular gas is dominated in 
circumnuclear regions in the luminous starburst mergers; i.e., 
$\Sigma$(gas) $\simeq \Sigma$(H$_2$).
Therefore, the above relationship suggests a star formation law; 
$\Sigma$(SFR) $\propto \Sigma$(gas).
We suggest that this star formation law favors the gravitational
instability scenario rather than the cloud-cloud collision one.
\end{abstract}




\section{INTRODUCTION}

Since the discovery of ultraluminous infrared galaxies (ULIGs:
Soifer et al. 1984; Wright, Joseph, \& Meikle
1984; see for a review Sanders \& Mirabel 1996),
much attention has been paid to this class of galaxies 
(hereafter starburst mergers\footnote{In this Letter,
we do not use a more popular terminology, ultraluminous infrared galaxies,
because our sample studied here contains mergers with
$L$(FIR) $ < 10^{12} L_\odot$.}). Most of these galaxies show morphological
evidence for galaxy mergers or strong galaxy interaction.
There is still a controversy on the origin of their huge
infrared luminosities; i.e., a pure starburst or  a hidden central engine of
active galactic nuclei (AGN). Recently, growing evidence
has been accumulated for the relative importance of 
starbursts in their central regions (e.g., Shaya et al. 1994; Genzel et al. 1998; 
Taniguchi, Trentham, \& Shioya 1998 and references therein).
However, an important question still remains; what controls the star formation process
in the starburst mergers ?
Therefore, it is very important to investigate this problem.

A number of previous CO observations have shown that the starburst mergers have 
a lot of molecular gas up to $\sim 10^{10} M_\odot$ (Young et al. 1986a, 1986b;
Sanders et al. 1988; Solomon \& Sage 1988; Sanders, Scoville, \& Soifer 1991).
Moreover, the molecular gas is usually distributed in the inner 1 kpc region
in these mergers [Scoville et al. 1991 (hereafter S91); Scoville, Yun, \& Bryant 1997; 
Downes \& Solomon 1998 (hereafter DS98)]. Therefore, it has been considered that 
the accumulation of a lot of molecular gas toward the central region
can lead to the intense starbursts in the mergers. Indeed,
numerical simulations have shown that galaxy mergers can fuel gas clouds toward the
central region of the merger products (Barnes \& Hernquist 1991; Mihos \& Hernquist 1994).
Here a question arises as how the intense starbursts were triggered in the fueled gas.
There may be three alternative basic physical processes  to cause the intense 
starbursts; 1) the gravitational 
instability in nuclear gas disks (e.g., Kennicutt 1998 and references therein),
2) efficient cloud-cloud collision driven by violent gas motion in their central 
regions (e.g., S91), and 3) triggering through the dynamical 
disturbance driven by supermassive binary black holes (Taniguchi \& Wada 1996). 
In order to investigate which process is most important in the starburst mergers,
we analyze two data sets of high-luminosity starburst mergers
mapped in CO($J$=1--0) using the Owens Valley 
Radio Observatory (OVRO) interferometer (S91) and the IRAM one (DS98).


\section{DATA}

The first data set is taken from S91 who 
presented a summary of the OVRO CO($J$=1--0) mapping
of 18 high-luminosity infrared galaxies.
Almost all the galaxies show evidence either for galaxy mergers or 
for strong galaxy interaction.
This sample covers a wide range of far-infrared (FIR) luminosities from 
$\sim 10^{10} L_\odot$ to $\sim 10^{12} L_\odot$.
The second data set is taken from DS98 who performed
detailed CO($J$=1--0) and CO($J$=2--1) observations of ten ULIGs.
Though four ULIGs among this sample are also included in the OVRO sample
(Arp 220, Mrk 231, VII Zw 31, and IRAS 17208$-$0014), we deal with this IRAM sample 
as an independent sample appreciating the two independent CO mapping 
programs using the two different radio interferometers.
Although these two samples are not statistically complete ones, 
they are highly useful in investigating a star formation law in starburst mergers.
In Table 1, we give a summary of observational properties of the 
individual galaxies of the two samples.
We use the distances of galaxies estimated using both recession velocities 
corrected to the Galactic Standard of Rest (GSR)
following the manner described in de Vaucouleurs et al. (1991) and 
a Hubble constant $H_0 = 75$ km s$^{-1}$ Mpc$^{-1}$. The heliocentric 
velocities are taken from de Vaucouleurs et al. (1991), S91, and DS98
(see Table 1). 

Molecular gas masses for all the galaxies in the two samples are derived 
using a conversion factor,
$\alpha = N_{\rm H_2}/I$(CO) = $3\times 10^{20}$ cm$^{-2}$ (K km s$^{-1}$)$^{-1}$
where $N_{\rm H_2}$ is the H$_2$ column density and $I$(CO) is the integrated 
intensity of the CO($J$=1--0) emission (S91).
The FIR data are compiled from the IRAS Faint Source Catalog
(Moshir et al. 1992).
The FIR luminosities are estimated using
$L$(FIR)$=4\pi D^2
1.26\times 10^{-11} [2.58\times S(60) + S(100)]$ (ergs s$^{-1}$)
where $S$(60) and $S$(100) are the IRAS 60 $\mu$m and 100 $\mu$m fluxes
in units of Jy and $D$ is the distance of galaxies in units of cm
(Helou, Soifer, \& Rowan-Robinson 1985). Finally $L$(FIR) is given in 
units of $L_\odot$.
In later analysis, in addition to $L$(FIR) and $M_{\rm H_2}$, 
we also use their surface densities; $\Sigma$(FIR) ($L_\odot$ pc$^{-2}$) and
$\Sigma$(H$_2$) ($M_\odot$ pc$^{-2}$), respectively.
These densities for each galaxy are normalized by a surface 
$A = \pi r^2$ where $r$ is the radius in units of pc given in S91 and DS98.
Note that for the four ULIGs commonly observed by both S91 and DS98, the CO sizes
in DS98 are systematically smaller than those in S91 (see Table 1).
We use the CO sizes given in DS98 for these four ULIGs
in the later analysis.

\section{RESULTS}

In Figure 1, we show diagrams of log $L$(FIR) ($L_\odot$), log $\Sigma$(FIR)
($L_\odot$ pc$^{-2}$), and  log $L$(FIR)/$M_{\rm H_2}$ ($L_\odot/M_\odot$)
as functions of log $\Sigma$(H$_2$) ($M_\odot$ pc$^{-2}$) and
log $S$(60)/$S$(100) for the two samples of S91 and DS98. 
Among the six relationships shown in Figure 1, we find the following 
two significant correlations for both the samples; 
a) log $\Sigma$(FIR) vs. log $\Sigma$(H$_2$), and
b) log $L$(FIR)/$M_{\rm dust}$ vs. log $S$(60)/$S$(100).

We obtain the nearly linear relationship between
$\Sigma$(FIR) and $\Sigma$(H$_2$); $\Sigma$(FIR) $\propto \Sigma$(H$_2$)$^{1.02\pm0.06}$
for the S91 sample, $\Sigma$(FIR) $\propto \Sigma$(H$_2$)$^{1.16\pm0.13}$
for the DS98 sample, and $\Sigma$(FIR) $\propto \Sigma$(H$_2$)$^{1.01\pm0.06}$
for the total (S91+DS98) sample. Note that the new data in DS98 are adopted 
for the four common ULIGs in the total sample. Therefore, the total number of ULIGs 
in the total sample is 24. The correlation coefficient for the total sample 
is quite high, $R \simeq 0.96$. 
Since $\Sigma$(FIR) can be regarded as the star formation rate (SFR) per 
unit surface area (e.g., Kennicutt 1998),
i.e., $\Sigma$(SFR), this correlation implies that the SFR is controlled
locally by the surface gas density.
On the other hand, since there is no significant correlation between $L$(FIR) and
$\Sigma$(H$_2$), the total FIR luminosity is governed not by the local
gas density but by the total gaseous content accumulated in the central region of 
starburst mergers. 

The correlation between $L$(FIR)/$M_{\rm H_2}$ and $S$(60)/$S$(100)
has been already found in a number of studies (e.g., Young et al. 1986a).
The correlation found here is not as tight as that for 
$\Sigma$(FIR) vs. log $\Sigma$(H$_2$). 
The data points of both NGC 4038/4039 and IRAS 00057+4021 make the 
correlation worse though we do not understand why these data are far 
from the correlation for the remaining galaxies.
Since the  $L$(FIR)/$M_{\rm H_2}$ ratio provides
a measure of star formation efficiency (Young et al. 1986a, 1986b; S91),
this correlation suggests that the higher star formation efficiency
leads to the production of more heating photons per dust grain,
resulting in the higher dust temperature.

Although S91 obtained a significant correlation between
log $L$(FIR)/$M_{\rm H_2}$ vs. log $\Sigma$(H$_2$) for 14 high-luminosity galaxies,
we do not find this correlation for all the galaxies in S91 and for the galaxies
in DS98. 
The little correlation between $L$(FIR)/$M_{\rm H_2}$ and $\Sigma$(H$_2$)
means that the star formation efficiency is not controlled by the surface
gas density. 

In Figure 2, we show diagrams of $\Sigma$(FIR) and $\Sigma$(H$_2$)
as a function of $A$ in logarithmic scales. 
It is shown that the two kinds of surface densities 
are almost inversely  proportional to the surface, $A$; 1)
$\Sigma$(FIR) $\propto A^a$: $a = -0.77\pm0.15$ for the S91 sample,
$-1.05\pm0.15$ for the DS98 sample, and $-0.85\pm0.13$ for the total sample,
and 2) $\Sigma$(H$_2$) $\propto A^b$: $b = -0.74\pm0.13$ for the S91 sample,
$-0.90\pm0.09$ for the DS98 sample, and $-0.82\pm0.12$ for the total sample.
The correlation coefficients for the total sample are $-0.80$ and $-0.82$,
respectively. 
Therefore, one important thing in the luminous starburst mergers is
how the gas clouds are accumulated in smaller regions.
We suggest that more concentrated gaseous systems tend to
experience more intense starbursts such that
$\Sigma$(SFR) $\propto \Sigma$(gas).


\section{DISCUSSION}

We consider the physical meaning of the linear relationship
between $\Sigma$(FIR) and $\Sigma$(H$_2$).
As mentioned before, this relationship suggests the star formation law
of  $\Sigma$(SFR) $\propto$  $\Sigma$(H$_2$).
This kind of relation has been discussed by many authors in terms of
so-called Schmidt law (Schmidt 1959); i.e., the star formation law remains
the simple gas-density power law; i.e., $\Sigma$(SFR) $\propto$  $\Sigma$(gas)$^N$.
In the derivation of the Schmidt law, it is recommended to use the total
(i.e., H {\sc i} + H$_2$) gas mass density. However, since the nuclear gas in
the high-luminosity starburst mergers is often dominated by molecular gas
(Mirabel \& Sanders 1989; see also Sanders \& Mirabel 1996), we can
neglect the contribution of H {\sc i} gas (Kennicutt 1998); 
i.e., $\Sigma$(gas) $\simeq \Sigma$(H$_2$).
Therefore, the linear correlation between $\Sigma$(FIR) and $\Sigma$(H$_2$)
suggests that the star formation in the luminous starburst mergers 
can be described by a Schmidt law with $N \simeq 1$.
 
For disks of normal galaxies, Kennicutt (1989) derived a relation of
$N \simeq 1.3\pm0.3$ (see also, Buat, 
Deharveng, \& Donas 1989; Buat 1992; Bosseli 1994; Deharveng et al. 1994).
Recently, Kennicutt (1998) derived a star formation law for a sample of
36 infrared-selected starburst galaxies; $N \simeq 1.40 \pm 0.13$.
This Schmidt law is much steeper than that derived in this study. 
This seems as due to the difference in sampling
between Kennicutt (1998) and ours. In his sample, a number of less luminous
starburst galaxies (e.g., IC 342, NGC 6946, and so on) are included. Although these
galaxies experience the moderate star forming activity, they are basically
isolated ordinary spiral galaxies. On the other hand, our sample comprises mostly
major mergers.
In fact, the steeper Schmidt law derived by Kennicutt (1998) appears 
to be biased by the inclusion of the less-luminous, non-merger galaxies.

As mentioned in Section 1, 
there are three alternative mechanisms to initiate starbursts
in merging galaxies; 1) the gravitational
instability in nuclear gas disks (e.g., Kennicutt 1998 and references therein),
2) efficient cloud-cloud collision driven by violent gas motion in their central
regions (e.g., Young et al. 1986a; S91), 
and 3) triggering through the dynamical
disturbance driven by a supermassive binary (Taniguchi \& Wada 1996).
A Schmidt law with $N \simeq 2$ can be achieved by the star formation driven by
cloud-cloud collisions (Scoville, Sanders, \& Clemens 1986).
On the other hand,  a Schmidt law with $N \simeq 1$ -- 1.5 implies the star
formation triggered by the gravitational instability of gaseous disks
(Kennicutt 1989, 1998; Larson 1988; Elmegreen 1994).
The Schmidt law derived here (i.e., $N \simeq 1$) favors the gravitational 
instability scenario rather than the cloud-cloud collision model. 
Finally, we cannot rule out the third possibility because this model also adopts 
the gravitational instability formalism.

Recently, Taniguchi, Trentham, \& Shioya (1998)
proposed a starburst-driven starburst mechanism
in order to understand the formation of blue super star clusters
in ULIGs such as Arp 220 (see also Shioya, Taniguchi, \& Trentham 1998).
In such a mechanism the infalling dense gas disk
is unstable gravitationally and collapses to form massive gaseous clumps.
These clumps are exposed
to the external high pressure from the superwind\footnote{A blast wave
driven by a collective effect of a large number of supernovae
in the very core of the galaxy (e.g., Heckman, Armus, \& Miley 1990;
Heckman et al. 1996).}
driven by the ongoing starburst at the very center of the galaxy.
This external pressure leads to
further collapse of the clumps which in turn leads to massive star
formation in them. Our result is consistent with this scenario.

Starbursts in major mergers have been also studied by numerical simulations
(e.g., Mihos \& Hernquist 1994 and references therein). In most the simulations with
an SPH code, the star formation law is often assumed to follow a Schmidt law.
For example, Mihos \& Hernquist (1994) adopted the Schmidt law with $N = 1.5$. 
However, our result shows that the Schmidt law with $N \simeq 1$ 
describes the star formation in the actual high-luminosity major mergers.
Therefore, we recommend that future numerical simulations for 
gas-rich major mergers will use the Schmidt law with $N \simeq 1$.

\vspace{0.5cm}

We would like to thank an anonymous referee for useful comments.
Y.O. was supported by the Grant-in-Aid for JSPS Fellows by
the Ministry of Education, Science, Sports and Culture.
This work was supported in part by the Ministry of Education, Science,
Sports and Culture in Japan under Grant Nos. 07055044, 10044052, and 10304013.


\newpage


\begin{deluxetable}{lcccccccccc}
\tablewidth{36pc}
\tablenum{1}
\tablecaption{Properties of the sample of galaxies}
\tablehead{
\colhead{(1)} &
\colhead{(2)} &
\colhead{(3)} &
\colhead{(4)} &
\colhead{(5)} &
\colhead{(6)} &
\colhead{(7)} &
\colhead{(8)} &
\colhead{(9)} &
\colhead{(10)} &
\colhead{(11)}
}
\startdata
\multicolumn{11}{l}{Scoville et al. (1991) sample}\nl
\tableline
Mrk231 & 165$^b$ & $<$2.76 & 32.0 & 30.3 & 1.06 & 12.08 & 10.51 & 1.57 & $>$4.71 & $>$3.13 \nl
I17208\tablenotemark{a} & 172$^d$ & 1.28 & 31.1 & 34.9 & 0.89 & 12.13 & 10.73 & 1.40 & 5.42 & 4.02 \nl
Arp220 & 74$^b$ & 0.29 & 103.8 & 112.4 & 0.92 & 11.91 & 10.17 & 1.73 & 6.50 & 4.76 \nl
VIIZw31 & 219$^d$ & 2.67 & 5.6 & 9.6 & 0.58 & 11.66 & 10.46 & 1.20 & 4.30 & 3.11 \nl
I10173\tablenotemark{a} & 194$^c$ & 3.27 & 5.8 & 5.5 & 1.06 & 11.48 & 9.95 & 1.53 & 3.96 & 2.42 \nl
N6240 & 98$^b$ & 1.65 & 22.7 & 27.8 & 0.82 & 11.52 & 10.03 & 1.49 & 4.58 & 3.09 \nl
IC694 & 42$^c$ & 0.30 & 103.7 & 107.4 & 0.97 & 11.41 & 9.59 & 1.82 & 5.96 & 4.14 \nl
VV114 & 81$^c$ & 0.93 & 22.6 & 30.4 & 0.74 & 11.35 & 10.03 & 1.33 & 4.92 & 3.59 \nl
N1614 & 62$^b$ & 0.60 & 32.3 & 32.7 & 0.99 & 11.25 & 9.78 & 1.47 & 5.19 & 3.72 \nl
Arp55 & 160$^c$ & 3.13 & 6.0 & 10.2 & 0.58 & 11.41 & 10.22 & 1.19 & 3.92 & 2.73 \nl
N1068 & 14$^b$ & 0.10 & 176.2 & 224.0 & 0.79 & 10.74 & 9.46 & 1.28 & 6.21 & 4.93 \nl
N7469 & 67$^b$ & 0.82 & 25.9 & 34.9 & 0.74 & 11.26 & 9.89 & 1.37 & 4.94 & 3.57 \nl
Zw049.057 & 52$^c$ & 0.40  & 20.8  & 29.4 & 0.71 & 10.95 & 9.67 & 1.28  & 5.24 & 3.96 \nl
N828  & 73$^b$ & 0.92 & 10.9 & 22.8 & 0.48 & 11.03 & 10.09 & 0.94 & 4.61 & 3.67 \nl
N2146 & 17$^b$ & 0.33 & 131.0 & 184.2 & 0.71 & 10.78 & 9.43 & 1.35 & 5.26 & 3.90 \nl
N3079 & 20$^b$ & 0.26 & 44.5 & 89.2 & 0.50 & 10.52 & 9.56 & 0.97 & 5.21 & 4.25 \nl
N520  & 28$^b$ & 0.38 & 30.9 & 45.8 & 0.67 & 10.58 & 9.47 & 1.11 & 4.92 & 3.80 \nl
N4038/9 & 26$^b$ & 0.49 & 39.5 & 72.3  & 0.55  & 10.65 & 9.07  & 1.58  & 4.78  & 3.20 \nl
\nl
\multicolumn{11}{l}{Downes \& Solomon (1998) sample}\nl
\tableline
I00057\tablenotemark{a} & 181$^d$ & 0.48 & 4.5 & 4.3 & 1.04 & 11.31 & 10.27 & 1.04 & 5.44 & 4.40 \nl
I02483\tablenotemark{a} & 207$^d$ & 0.90 & 4.0 & 6.9 & 0.58 & 11.46 & 10.16 & 1.31 & 5.06 & 3.75 \nl
VIIZw31 & 219$^d$ & 1.27 & 5.6 & 9.6 & 0.58 & 11.66 & 10.70 & 0.96 & 4.95 & 3.99 \nl
I10565\tablenotemark{a} & 172$^d$ & 0.67 & 12.1 & 15.1 & 0.80 & 11.73 & 10.38 & 1.35 & 5.59 & 4.24 \nl
Mrk231 & 165$^b$ & 0.36 & 32.0 & 30.3 & 1.06 & 12.08 & 10.35 & 1.73 & 6.47 & 4.74 \nl
Arp193 & 92$^b$ & 0.25 & 15.4 & 25.2 & 0.61 & 11.34 & 10.22 & 1.12 & 6.06 & 4.94 \nl
Mrk273 & 152$^b$ & 0.13 & 21.7 & 21.4 & 1.02 & 11.85 & 10.33 & 1.51 & 7.13 & 5.62 \nl
Arp220 & 74$^b$ & 0.16  & 103.8 & 112.4 & 0.92 & 11.91 & 10.43 & 1.48 & 7.00 & 5.52 \nl
I17208\tablenotemark{a} & 172$^d$ & 0.75 & 31.1 & 34.9 & 0.89 & 12.13 & 10.67 & 1.46 & 5.88 & 4.42 \nl
I23365\tablenotemark{a} & 260$^d$ & 0.63 & 7.1 & 8.4 & 0.85 & 11.85 & 10.39 & 1.46 & 5.75 & 4.29 \nl
\enddata
\tablecomments{
(1) Name,
(2) Distance (Mpc); heliocentric velocities are taken from (b) de Vaucouleurs 
et al. (1991), (c) S91, and (d) DS98,
(3) Radius (kpc),
(4) $S$(60) (Jy),
(5) $S$(100) (Jy),
(6) $S$(60)/$S$(100),
(7) log $L$(FIR) ($L_{\odot}$),
(8) log $M$(H$_{\rm 2}$) ($M_{\odot}$),
(9) log $L$(FIR)/$M$(H$_{\rm 2}$) ($L_{\odot}$/$M_{\odot}$),
(10) log $\Sigma$(FIR) ($L_{\odot}$/pc$^2$), and
(11) log $\Sigma$(H$_2$) ($M_{\odot}$/pc$^2$)
}
\tablenotetext{a}{I17208 = IRAS 17208$-$0014; I10173 = IRAS 10173+0828; Zw049 = Zw049.057;
I00057 = IRAS 00057+4021; I02483 = IRAS 02483+4302; I10565 = IRAS 10565+2448;
I23365 = IRAS 23365+3604.}
\end{deluxetable}


\figcaption{
Diagrams of log $L$(FIR), log $\Sigma$(FIR), and log $L$(FIR)/$M_{\rm H_2}$ 
as functions of log $\Sigma$(H$_2$) and log $S$(60)/$S$(100).
The OVRO sample is shown by filled circles while the IRAM one is
shown by open circles. The filled squares for the four ULIGs observed by S91
(Arp 220, Mrk 231, VII Zw 31, and IRAS 17208$-$0014) show the data points
when we adopt the sizes obtained by DS98. However, note that the data points
of the four ULIGs in the log $\Sigma$(FIR) vs. log $S$(60)/$S$(100) are the same as
those for DS98.
\label{fig1}
}

\figcaption{
Diagrams of log $\Sigma$(FIR) and log $\Sigma$(H$_2$)
as a function of the surface $A$.
The symbols have the same meanings as those in Figure 1.
\label{fig2}
}

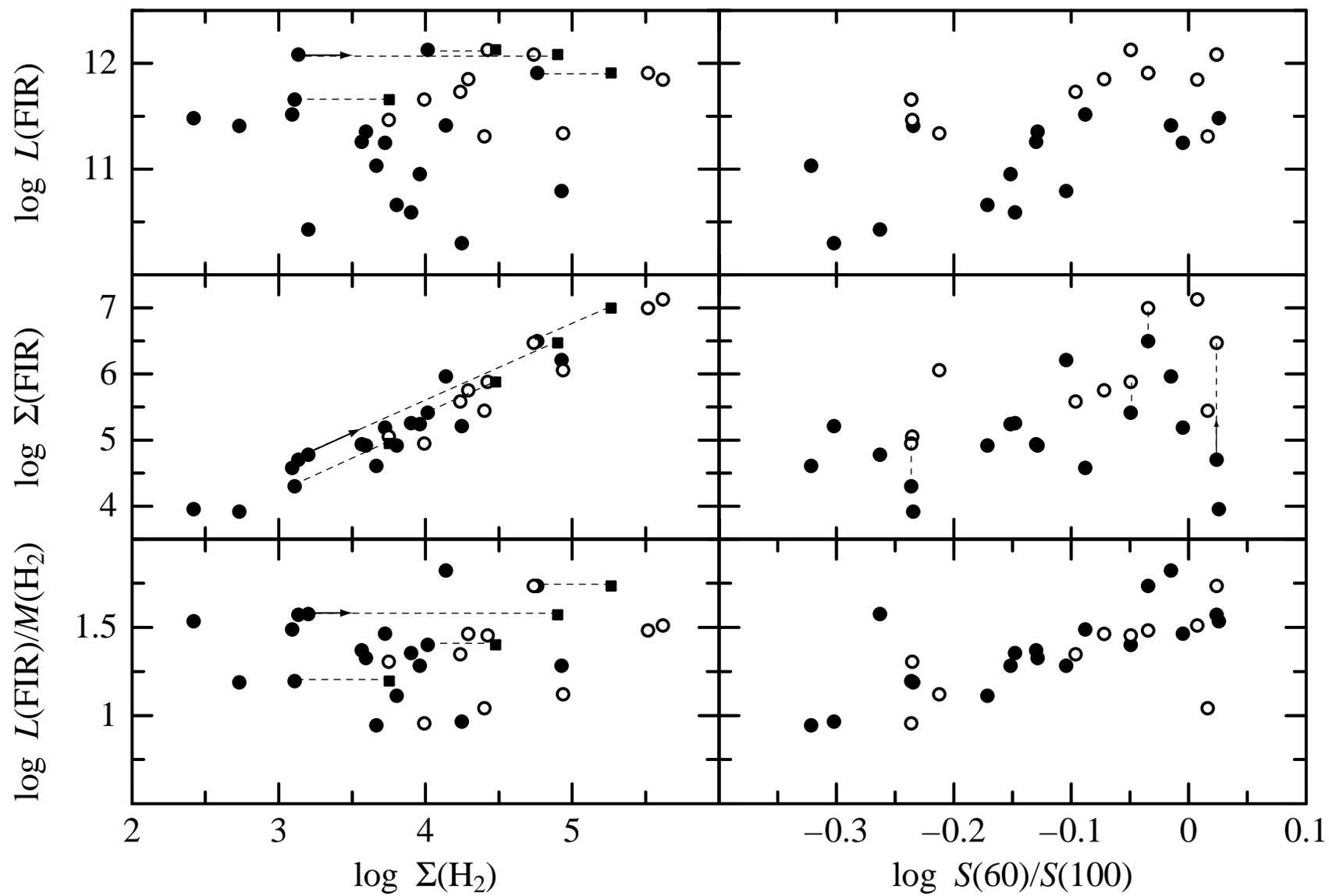
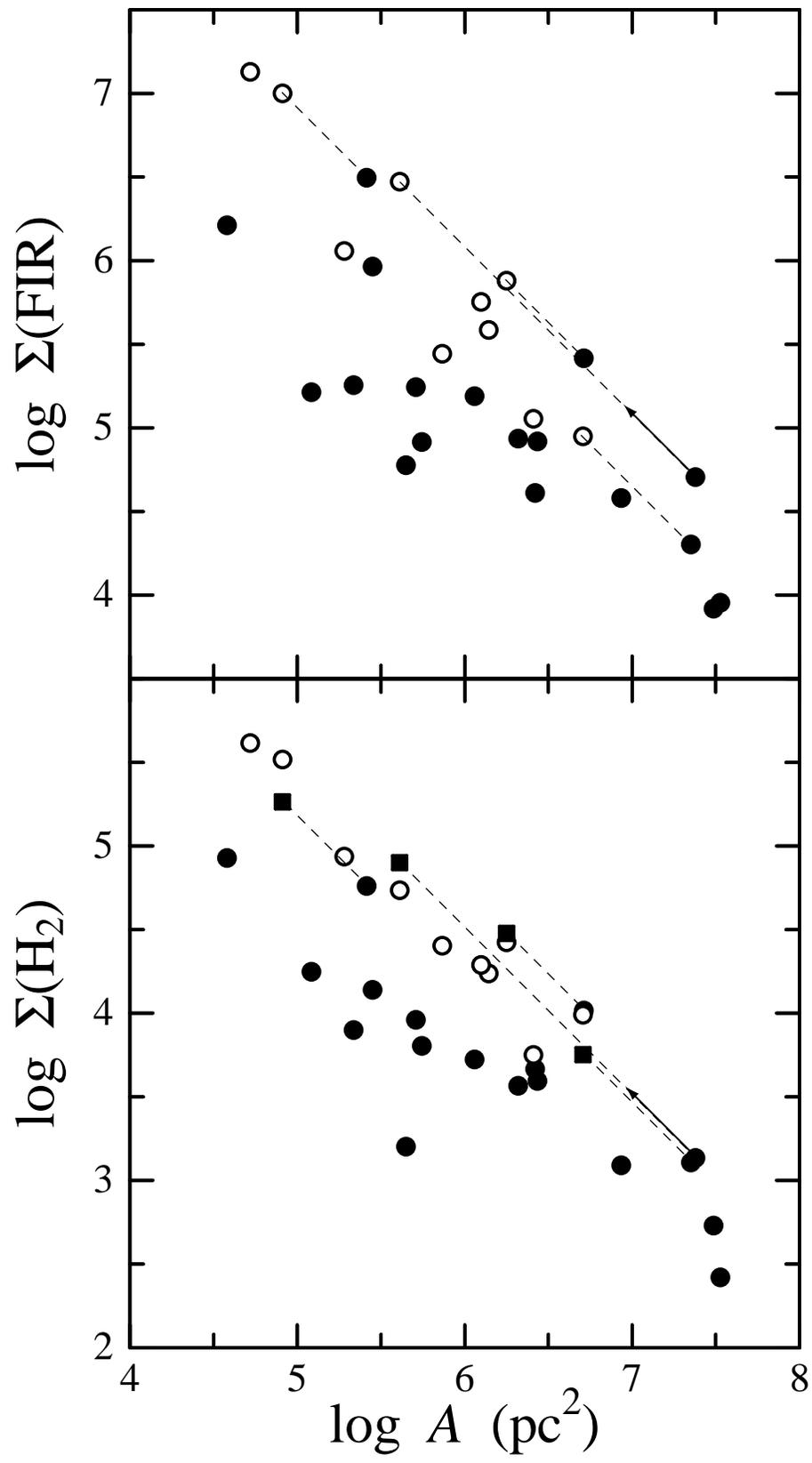
\end{document}